\documentclass[a4paper,12pt]{article}
\usepackage{blkarray}
\usepackage{amsfonts}
\usepackage{footmisc}
\usepackage{amsmath}
\usepackage{setspace}
\usepackage[top=1.5in, bottom=2in, left=1.2in, right=1.2in]{geometry}
\usepackage{graphicx}
\usepackage{amssymb}
\usepackage{fancyhdr}
\providecommand{\U}[1]{\protect\rule{.1in}{.1in}}
\newtheorem{theorem}{Theorem}

\newtheorem{definition}[theorem]{Definition}

\newenvironment{proof}[1][Proof]{\noindent\textbf{#1.} }{\ \rule{0.5em}{0.5em}}

\def\fr{\frac}
\def\be{\begin{equation}}
\def\ee{\end{equation}}
\def\ba{\begin{eqnarray}}
\def\ea{\end{eqnarray}}
\def\pa{\partial}
\def\ra{\rightarrow}
\def\na{\nabla}
\def\s{\sigma}
\def\l{\lambda}
\def\e{\varepsilon}
\def\a{\alpha}
\def\b{\beta}
\def\g{\gamma}
\def\d{\delta}
\def\t{\tau}

\def\pt{\phantom{a}}
\def\w{\omega}

\newcommand{\captionfonts}{\normalsize}

\makeatletter  
\long\def\@makecaption#1#2{%
  \vskip\abovecaptionskip
  \sbox\@tempboxa{{\captionfonts #1: #2}}%
  \ifdim \wd\@tempboxa >\hsize
    {\captionfonts #1: #2\par}
  \else
    \hbox to\hsize{\hfil\box\@tempboxa\hfil}%
  \fi
  \vskip\belowcaptionskip}
\makeatother   

\begin{document}

\title{AFT Gravitational Model \\ Unity of All Elementary Particles in $Sp(12,C)$}
\author{Diego Marin}
\date{}

\maketitle
\normalsize
\begin{abstract}\begin{spacing}{1.2}
A new unifying theory was recently proposed
in the publication \emph{Arrangement field theory - beyond strings and
loop gravity -}\cite{Arrangement}. Such theory describes all fields (gravitational,
gauge and matter fields) as entries in a matricial superfield which
transforms in the adjoint representation of $Sp(12,C)$. In this paper
we show how this superfield is built and we introduce a new mechanism
of symmetry breaking which doesn't need Higgs bosons.
\end{spacing}
\end{abstract}

\newpage
\setcounter{page}{1}
\tableofcontents

\newpage

\section{Introduction}

A new unifying theory was recently proposed
in the publication \emph{Arrangement field theory - beyond strings and
loop gravity -}, edited by LAMBERT Academic Publishing\cite{Arrangement}.

Such theory describes gravitational, gauge and matter fields by
means of probabilistic spin-networks, ie collections of vertices
and edges where the existence of any edge is regulated by a quantum
amplitude. The best result of this approach is the manifestation
of gravity as a fictitious force which appears when a probabilistic spin
network is substituted by a medium state with fixed edges.

In this way the tetrad $e^\mu$ is not a dynamical field but an
appropriate function or distribution. Conversely, the $SO(1,3)$
connections remain dynamical fields as the other Yang-Mills fields.
However these define only a subensemble in the ensemble of $Sp(12,C)$
connections.

This group spontaneously appears in Arrangement Field Theory as
a consequence of its basilar assumptions, the same which correctly
predict Black Hole entropy.

Tangent space assumes $SO(1, 3)$ symmetry only when
gravity decouples from other forces. At that point also the real
space-time can obtain the same symmetry. This fact is coherent with
\emph{no-go theorem} of Coleman-Mandula \cite{nogo}, under which
\lq\lq $S$-matrix is Lorentz invariant if and only if the action
symmetry is $SO(1, 3) \otimes$ \emph{internal symmetries}''.

We start in section \ref{ricci} by constructing the Ricci scalar as
the (totally contracted) antisymmetrized second covariant derivative
of $Sp(12,C)$. We show that the only connections which contribute to
this term are the $SO(1,3)$ connections.

In section \ref{kinetic} we construct a kinetic term for $Sp(12,C)$
gauge fields. We extract the gravitational contribute, showing that
it reproduces the topological term of Gauss-Bonnet, so that the
motion equations result unchanged.

In section \ref{standard} we embed Standard Model symmetry
($SU(3) \otimes SU(2) \otimes U(1)$) and gravitational gauge
symmetry ($SO(1,3)$) inside a larger $Sp(12,C)$ symmetry.

Hence we assemble fermionic fields in such a way to fill up
the adjoint representation of this group. Doing this we discover
an approximate (global) flavour symmetry $SU(3) \otimes SU(3)$.

In section \ref{superfield} we combine bosonic and fermionic
fields in a unique superfield without need for new unseen particles.

In the last section we explicitly show the strangest prediction
of theory, ie the possibility to obtain an antigravitational force
by means of electromagnetic fields or other Yang-Mills fields.

\section{Ricci scalar}
\label{ricci}

In this section we define Ricci scalar in a modified
Palatini formalism which makes it suitable to describe
gravity as a branch of an unified force.

To do this we need to introduce two little known extensions
of complex numbers (sometimes called hyper-complex numbers)
that are \lq\lq Quaternions ($\mathbf{H}$)'' and \lq\lq
Hyperions ($\mathbf{Y}$)''.

\subsection{Quaternions}

We start by considering the ensemble of quaternions ($\mathbf{H}$), an associative normed division algebra over the real numbers. Such algebra was introduced by Hamilton in 1843\cite{Hamilton} and it's completely defined by relations:

$$ij = -ji = k \qquad jk = -kj = i \qquad ki = -ik =j$$
$$i^2 = j^2 = k^2 = -1$$

\noindent The base elements $i,j,k$ satisfy the same algebra of Pauli matrices and thus they are good to describe rotations in the euclidean three-dimensional space. We think about them as imaginary unities, so that a generic quaternion $q$ takes the form

$$q = a + ib + jc + kf \qquad \pt\text{with}\pt a,b,c,f \in \mathbf{R}.$$

\noindent Pay attention that, dislike complex algebra, quaternionic algebra isn't commutative (in general $pq \neq qp$).

\subsection{Hyperions}

We define an extension of $\mathbf{H}$ by introducing
a new imaginary unit $I$ which satisfies

$$I^2 = -1 \qquad I^\dag = -I$$
$$[I,i] = [I,j] = [I,k] = 0$$

\noindent In this way a generic number assumes the form

$$v = a + Ib+ ic + jd + ke + iIf+ jIg + kIh, \qquad a,b,c,d,e,f,g,h \in \mathbf{R}$$
$$v = p + Iq, \qquad p,q \in \mathbf{R}$$

\noindent We call this numbers \lq\lq Hyperions'' and we indicate their ensemble with $Y$.
It's easy to see their correspondence with even
products of Gamma matrices, explicitly

$$1 \Leftrightarrow \g_0 \g_0 = 1 \qquad I \Leftrightarrow \g_5 = \g_0 \g_1 \g_2 \g_3$$
$$i \Leftrightarrow \g_2 \g_1 \qquad iI \Leftrightarrow \g_0 \g_3$$
$$j \Leftrightarrow \g_1 \g_3 \qquad jI \Leftrightarrow \g_0 \g_2$$
$$k \Leftrightarrow \g_3 \g_2 \qquad kI \Leftrightarrow \g_0 \g_1$$

\noindent Note that imaginary units $i,j,k,iI,jI,kI$ satisfy the Lorentz algebra,
with $i,j,k$ which describe rotations and $iI, jI, kI$ which describe boosts.

\begin{definition}[bar-conjugation]
The bar-conjugation is an operation which exchanges $I$ with $-I$ (or $\g_0$ with
$-\g_0$ in the $\g\g$-representa\-tion). Explicitly, if $v = a + Ib+ ic + jd + ke + iIf+
jIg + kIh$ with $a,b,c,d,e,f,g,h \in \mathbf{R}$, then $\bar{v} = a - Ib+ ic + jd +
ke - iIf - jIg - kIh$.
\end{definition}

\begin{definition}[pre-norm]
The pre-norm is a complex number with $I$ as imaginary unit (we say \lq\lq I-complex number'').
Given an hy\-pe\-rion $v$, its pre-norm is $|v| = (\bar{v}^\dag v)^{1/2}$. If $v \in \mathbf{H}$,
its pre-norm coincides with usual norm $(v^\dag v)^{1/2}$.
\end{definition}

\noindent Note that every hyperion $v$ can be written in the polar form

$$v = |v|e^{ia+jb+kc+iId+jIe+kIf} \qquad a,b,c,d,e,f$$
$$|v|^2 = \bar{v}^\dag v = |v|e^{-(ia+jb+kc+iId+jIe+kIf)} |v|e^{ia+jb+kc+iId+jIe+kIf}= |v|^2.$$

\noindent Moreover, the norm of any hyperion $v$ is the norm of its prenorm, indicated with $||v||$.

\begin{definition}[Hyper-unitary matrices] A square matrix $U$ with elements in $\mathbf{C}$ is called unitary matrix if it satisfies

$$U^\dag U = U U^\dag = 1.$$
Unitary matrices $n \times n$ define a Lie group $U(n)$ having real dimensions $n^2$.
Similarly, a square matrix $U$ with elements in $\mathbf{H}$ is called hyper-unitary if
$$U^\dag U = U U^\dag = 1.$$
Hyper unitary matrices with elements in $\mathbf{H}$ define a Lie group $Sp(n)$ having real dimensions $n(2n+1)$.
Finally, a square matrix $U$ with elements in $\mathbf{Y}$ is called hyper-unitary if
$$\overline{U}^\dag U = U \overline{U}^\dag = 1.$$
Hyper unitary$\pt$ matrices $\pt$with $\pt$elements in $\pt\mathbf{Y}$ define a Lie group

\noindent $Sp(2n,C)$ having real dimensions $2n(2n+1)$.
It's easy to see that $Sp(n)$ is the compact real form of $Sp(2n,C)$. As consequence, any generator $u$ in the $sp(n)$ algebra gives rise to a couple of generators $(u, Iu)$ in the $sp(2n, C)$ algebra. Moreover, $sp(2,C) \approx so(1,3)$ and $sp(1) \approx su(2) \approx so(3)$.

\end{definition}

\subsection{Ricci scalar with hyperions}

Given a gauge field $\w_\mu$ in $so(1,3)$ and a complex
tetrad $e^\mu$, we define

\be A_\mu = \w^{ab}_\mu \g_{a}\g_{b} \qquad\qquad h^{\mu\nu} = Re\,(e^{\dag \mu}_a e^\nu_b \eta^{ab})\label{gaugey} \ee
$$e^\mu = e^{\mu a} \g_0 \g_a \qquad\qquad \bar{e}^\mu = e^\mu (\g_0 \rightarrow-\g_0)$$
$$\Rightarrow \bar{e}^{\dag \mu} e^{\nu} = e^{\dag \mu a}e^{\nu b} \g_{a} \g_{b} \quad \Rightarrow h^{\mu\nu} = \fr 14 Re\,\left[ tr(\bar{e}^{\dag \mu} e^{\nu})\right] $$

\noindent Note that our definitions are the same to require $\bar{A}^\dag =-A$
in the hyperions framework. We claim that Ricci scalar can be written as

$$R(x) = -\fr 18 tr\left(\left(\pa_\mu A_\nu - \pa_\nu A_\mu + [A_\mu, A_\nu]\right) \bar{e}^{\dag\mu} e^\nu\right)$$

\noindent To verify our statement we expand first the commutator

\ba [A_\mu, A_\nu] &=& \w^{ab}_\mu \w^{cd}_\nu \left( \g_a\g_b\g_c\g_d - \g_c \g_d \g_a \g_b \right) \nonumber \\
                &=& \fr 12 \w^{ab}_\mu \w^{cd}_\nu \left( \g_a\{\g_b,\g_c\}\g_d - \g_c \{\g_d, \g_a\} \g_b \right)  + \nonumber \\
                && +\fr 1 {2}\w^{ab}_\mu \w^{cd}_\nu \left( \g_a [\g_b,\g_c ]\g_d - \g_c [\g_d, \g_a] \g_b \right)\nonumber \\
                &=& \left(\w^{ab}_\mu \w_{b\nu}^{\pt d} - \w^{ab}_\nu \w_{b\mu}^{\pt d} \right)\left( \g_a \g_d \right)  + \nonumber \\
                && +\fr 1 {4!}\w^{ab}_\mu \w^{cd}_\nu \left( \e_{abcd} \e^{efgh} \g_e \g_f \g_g \g_h \right)\nonumber \\
                &=& [\w_\mu, \w_\nu]^{ab}\g_a\g_b + \w^{ab}_\mu \w^{(D)}_{ab\nu} \,\g_5 \ea

\noindent In the last line we have defined $\w^{(D)}_{ab \nu} = \e_{abcd} \w_\nu^{cd}$. Hence

\ba R(x) &=& -\fr 18 tr(\g_a \g_b \g_c \g_d)\left( \pa_\mu \w^{ab}_\nu - \pa_\nu \w^{ab}_\mu + [\w_\mu,\w_\nu]^{ab}\right)e^{\dag c\mu} e^{d\nu} - \nonumber \\
        && -\fr 18 tr(\g_5 \g_b \g_c) \w^{ab}_\mu \w^{(D)}_{ab\nu} e^{\dag c\mu} e^{d\nu} \ea

\noindent Consider now the relations

$$\fr 14 tr(\g_a \g_b \g_c \g_d) = \eta_{ab}\eta_{cd} - \eta_{ac}\eta_{bd} + \eta_{ad}\eta_{bc}$$
$$tr(\g_5 \g_b \g_c) = 0$$

\noindent We obtain

$$R(x) = \left( \pa_\mu \w^{ab}_\nu - \pa_\nu \w^{ab}_\mu + [\w_\mu,\w_\nu]^{ab}\right)e^{\dag \mu}_a e^{\nu}_b $$
which is the usual definition.

We can move freely from matrices $\g$ to hyperions,
substituting $tr$ with $4$. In this way

\ba R(x) &=& -\fr 12 \left(\pa_\mu A_\nu - \pa_\nu A_\mu + [A_\mu, A_\nu]\right) \bar{d}^{\dag\mu} d^\nu \nonumber \\
        &=& -\fr 12 [\na_\mu,\na_\nu]\bar{e}^{\dag\mu} e^\nu \nonumber \\
        && \pt \nonumber \\
\na_\mu &=& \pa_\mu + A_\mu \qquad\qquad A_\mu, e^\mu \in \mathbf{Y}\nonumber \\
e^\mu_a &=& Re\,e^\mu_a + I\,Im\,e^\mu_a \nonumber \\
e^\mu &=& Re\,e^{\mu 0} + i I\, Re\,e^{\mu 3} + j I\, Re\,e^{\mu 2} + k I\, Re\,e^{\mu 1} + \nonumber \\
 && + I\,Im\,e^{\mu 0} - i\, Im\,e^{\mu 3} - j\, Im\,e^{\mu 2} - k\, Im\,e^{\mu 1} \nonumber \ea

\noindent By definition (\ref{gaugey}) we have $\bar{A}^\dag = -A$. Moreover, when it acts on reasonable Hilbert spaces, the operator $\pa^\dag$ is equal to $-\pa$. This implies $\overline{\na}^\dag_\mu = -\na_\mu$ and then

\ba R(x) &=& \fr 12 [\overline{\na}^\dag_\mu,\na_\nu]\bar{e}^{\dag\mu} e^\nu \nonumber \ea

\subsection{Ricci scalar in the new paradigm}

We now suppose that gravity gauge group $SO(1,3)$ is only a subgroup
in a bigger $Sp(12,C)$. Gauge field for $Sp(12,C)$ are $6 \times 6$
matrices $A^{ij}$ with entries in $Y$. The $SO(1,3)$ subgroup has
$6$ generators which are the complex unities $i,j,k,Ii,Ij,Ik$ in $tr(A^{ij}) = \sum_i A^{ii}$.

We verify that other fields don't contribute to the following generalized Hilbert Einstein lagrangian:

\ba L_{HE} &=& \fr 12 tr\,[\overline{\na}^\dag_\mu , \na_\nu ]\, \overline{e}^{\dag\mu} e^\nu \label{espansione} .\ea
Expanding the covariant derivatives we obtain

\ba L_{HE} &=& \fr 12 \sum_{i} \{ \pa^\dag_\mu A_\nu^{ii} -\pa_\nu \bar{A}_\mu^{\dag ii} + [\bar{A}^\dag_\mu, A_\nu]^{ii} \}\,\overline{e}^{\dag\mu} e^{\nu} \nonumber \\
&=& \fr 12 \{ \pa^\dag_\mu tr\,A_\nu -\pa_\nu tr\,\bar{A}^\dag_\mu +
[tr\,\bar{A}^\dag_\mu , tr\,A_\nu] + \nonumber \\
        && \qquad\qquad\qquad +\sum_{i,k\neq i}[\bar{A}_\mu^{\dag ik} A^{ki}_\nu - A_\nu^{ik} \bar{A}^{\dag ki}_\mu]\} \, \bar{e}^{\dag\mu}e^{\nu} \nonumber \ea

\noindent Note that $[\bar{A}^{\dag ii},A^{jj}]$ is equal to zero when $i \neq j$ and then

$$\sum_{a} [\tilde{A}^{^\dag ii}_\mu, A^{jj}_\nu] = \sum_{ij}[\bar{A}^{^\dag ii}_\mu, A^{jj}_\nu]= [tr\,\bar{A}^\dag_\mu, tr\,A_\nu].$$

\noindent For what follows we write $L_{HE} = \fr 12 \sum_{ij} R^{ij}_{\mu\nu} \d^{ij}\bar{e}^{\dag\mu}e^\nu $ with

\ba R^{ij}_{\mu\nu} &=& \d^{ij}\pa^\dag_\mu tr\,A_\nu -\d^{ij}\pa_\nu tr\,\bar{A}^\dag_\mu + [tr\,\bar{A}^\dag_\mu , tr\,A_\nu] + \nonumber \\
        && \qquad\qquad +\sum_{i,k\neq i,j\neq k}[\bar{A}_\mu^{\dag ik} A^{kj}_\nu - A_\nu^{ik} \bar{A}^{\dag kj}_\mu] .\nonumber \\ && \label{curvature}\ea
$R^{ij}_{\mu\nu}$ is thus a generalization of curvature tensor. Consider now any skew hermitian matrix $W_\mu$ with elements $W_\mu^{ij} = A_\mu^{ij}$ for $i \neq j$ and $W_\mu^{ij} = 0$
for $i = j$. It belongs to the subalgebra of $sp(12,C)$ made by all null track generators.
This means that commutators between null track generators are null track generators too. In this way

$$ \sum_{i,j\neq i} [\bar{A}^{\dag ij}_\mu , A_\nu^{ji}] = tr [\bar{W}^\dag_\mu, W_\nu] = 0 .$$

\noindent Hence we can delete the mixed term in $L_{EH}$.

\ba L_{HE} &=& \fr 12 \{ \pa^\dag_\mu tr\,A_\nu -\pa_\nu tr\,\bar{A}^\dag_\mu +
               [tr\,\bar{A}^\dag_\mu , tr\,A_\nu]\}\bar{e}^{\dag\mu}e^{\nu}\nonumber \\
               &=& \!\! -\fr 12 [\overset{G}{\na}_\mu,\overset{G}{\na}_\nu]\bar{e}^{\dag\mu}(x^a)e^{\nu}(x^a) \nonumber \\
           &=& R .\ea

\noindent Here $\overset{G}{\na} = -\overset{G}{\overline{\na}^\dag}$ is the gravitational covariant
derivative $\overset{G}{\na} = \pa + tr\,A$. As we have claimed, we see that gauge fields in $R$ are only
the diagonal ones. Conversely, we'll show that other gauge fields in the Standard Model correspond
to non diagonal components.

\section{The kinetic term}
\label{kinetic}
Until now we have obtained no terms which describe gauge interactions.
In this section we find a such term, with the condition that it hasn't
to change Einstein equations. One option is as follows:

\ba
L_{GB}  &=& -tr\,\left[ \bar{\na}^\dag_\mu, \na_\nu\right] \bar{e}^{\dag\mu} e^\nu \left[\bar{\na}^\dag_\a, \na_\b \right] \bar{e}^{\dag\a} e^\b   \ea

\noindent We use newly the correspondence between
$(1,I,i,j,k,iI,jI,kI)$ and gamma matrices:

\ba L_{GB}  &=& -\fr 14 tr(\g_a\g_b\g_c\g_f \g_g \g_h \g_m \g_n)\cdot \nonumber \\
&& \qquad \cdot \left[ \bar{\na}^\dag_\mu, \na_\nu\right]^{ab} \bar{e}^{\dag c \mu} e^{f \nu} \left[\bar{\na}^\dag_\a, \na_\b \right]^{gh} \bar{e}^{\dag m \a} e^{n \b}  \nonumber \ea

\noindent Here we use letters $a,b,c,d$ for indices which run on Gamma matrices, $\a,\b,\mu,\nu$
for spatial coordinates indices and $ijk$ for gauge indices.

In the next section we'll see that physical fields arise in three families, determined by the
choice of a subspace inside $Y$. This is true both for fermionic and bosonic
fields. Thus the indices with letters $a,b,c,d$ run over the three families. Exploiting calculation

\ba L_{GB} &=& R^{ij}_{ab\mu\b} R^{ab ji}_{\nu\a} \bar{e}^{\dag\mu}_c e^{c\nu} \bar{e}^{\dag \a}_d e^{d\b} -
4R^{ij}_{ac\mu\b} \bar{e}^{\dag a\mu} R^{cb ji}_{\nu\a} e^{\a}_b e^{d\b} \bar{e}^{\dag \a}_d  + \nonumber   \\
        && + R^{ij}_{ac\mu\b} \bar{e}^{\dag a\mu} e^{c\b} R^{cb ji}_{\nu\a} \bar{e}^{\dag \nu}_c e^{\a}_b   \nonumber \\
        &=& R^{ij}_{ab\mu\b} R^{ab ji \mu\b} - 4 R^{ij}_{c \b} R^{c ji \b} + R^{ij}R^{ji} \label{quartic}\ea

\noindent $R^{ij}_{\b\mu}$ was defined in (\ref{curvature}), while $R^{ij}_\mu
= R^{ij}_{\b\mu} e^{\b}$ and $R^{ij} = R^{ij}_{\b\mu} e^{\b} e^{\dag\mu}$.
You understand in a moment that for $i \neq j$ we have $R^{ij}_{ac\b\mu}
R^{ji ac \b\mu} = tr\,\sum_{(ac)} F^{(ac)}_{\mu\nu} F^{(ac) \mu\nu}$.
The index $(ac)$ runs over three fields families and $F_{(ac)\mu\nu}$
is a strength field tensor. In this way the terms $R^{ij \nu}_\b R^{ji \b}_\nu$
and $R^{ij}R_{ji}$ are terms which mix families. Conversely, for $i=j$ we have

$$L_{GB} = R_{ac\b\mu} R^{ac\b\mu} + R^2 - 4 R^{\a}_\mu R^{\mu}_\a $$

which is the Gauss-Bonnet topological term and so it doesn't change the Einstein equations.

\subsection{Symmetry breaking}

The combination of $L_{HE}$ and $L_{GB}$ gives to gravitational
gauge field $\overset{G}{A}$ a potential with form

$$\overset{G}{A^2} - \overset{G}{A^4}.$$
This potential has non trivial minimums which imply a non-trivial
expectation value for $\overset{G}{A}$. Moreover, inside $S_{GB}$ we
find the following kind of terms for other fields $A$:

$$\langle \overset{G}{A^2} \rangle A^2 - A^4.$$
In this way we have a mass for gauge fields $A$ and another
potential with non-trivial minimums. Therefore, also gauge fields
$A$ have non-trivial expectation values. Finally, such expectation values
give mass to fermionic fields via terms

$$\psi^\dag \langle A \rangle \psi.$$
There is no need for a scalar Higgs boson. Obviously, inside $\langle \cdot \rangle$
there must be a contraction with $e^\mu$ to preserve covariance.

\section{Standard model interactions}
\label{standard}

In this section we construct a local field theory with gauge group $Sp(12,C)$,
showing that it includes gravitational field, $SU(5)$-Yang-Mills fields and
three families of fermions with local symmetry $SU(5)$.
A Grand Unified Theory based on $SU(5)$ symmetry
was already proposed by Howard Georgi and Sheldon Glashow
in 1974. To understand how this theory includes in turn the
Standard Model, please refer to the original work\cite{Georgi}.
Nevertheless our framework uses a very different mechanism
of symmetry-breaking which doesn't make use of Higgs bosons.
In this manner it circumvents the major problem in G-G model.
Such model predicts in fact the proton decay via virtual
Higgs bosons, a phenomenon never observed.

The gauge fields $A^{ij}$ are $6 \times 6$ skew adjoint hyperionic matrices $\bar{A}^\dag = -A$. These
matrices form the $Sp(12,\mathbf{C})$ algebra which has $156$ generators $\w$ with $\bar{\w}^\dag = -\w$.

$$\w = \left( \begin{array}[c]{cccccc}
 \vec{y}   & b+\vec{b}  & c+\vec{c}  & d+\vec{d}  & e+\vec{e}  & m+\vec{m}  \\
 -b+\vec{b}& \vec{a}_1  & f+\vec{f}  & g+\vec{g}  & h+\vec{h}  & p+\vec{p}  \\
 -c+\vec{c}& -f+\vec{f} & \vec{a}_2  & s+\vec{s}  & q+\vec{q}  & r+\vec{r}  \\
 -d+\vec{d}& -g+\vec{g} & -s+\vec{s} & \vec{a}_3  & k+\vec{k}  & t+\vec{t}  \\
 -e+\vec{e}& -h+\vec{h} & -q+\vec{q} & -k+\vec{k} & \vec{a}_4  & v+\vec{v}  \\
 -m+\vec{m}& -p+\vec{p} & -r+\vec{r} & -t+\vec{t} & -v+\vec{v} & \vec{a}_5  \\
\end{array} \right) $$

\noindent Consider now the subalgebra of the following form with complex
(not hyperionic) components except for $y$ which remains hyperionic:

$$\w = \left( \begin{array}[c]{cccccc}
 \vec{y}   & 0          & 0          & 0          & 0          & 0          \\
 0         & \vec{a}_1  & f+\vec{f}  & g+\vec{g}  & h+\vec{h}  & p+\vec{p}  \\
 0         & -f+\vec{f} & \vec{a}_2  & s+\vec{s}  & q+\vec{q}  & r+\vec{r}  \\
 0         & -g+\vec{g} & -s+\vec{s} & \vec{a}_3  & k+\vec{k}  & t+\vec{t}  \\
 0         & -h+\vec{h} & -q+\vec{q} & -k+\vec{k} & \vec{a}_4  & v+\vec{v}  \\
 0         & -p+\vec{p} & -r+\vec{r} & -t+\vec{t} & -v+\vec{v} & \vec{a}_5  \\
\end{array} \right) $$

\noindent Moreover we put the additional condition $\vec{a} = \sum_l \vec{a}_l = 0$.
The field $y = tr\,\w$ is the only one which contributes to Ricci scalar.
Conversely, all other fields belong to a $SU(5)$ subgroup, which defines the Georgi - Glashow
grand unification theory. The symmetry breaking in Georgi - Glashow model is induced by Higgs
bosons in representations which contain triplets of color. These color triplet Higgs can mediate
a proton decay that is suppressed by only two powers of GUT scale. However, our mechanism of
symmetry breaking doesn't use such Higgs bosons, but descends from the expectation values of
quadratic terms $AA$, which derive from non trivial minimums of a potential $AA - AAAA$. So we
circumvent the problem.

Restrict now the attention to the $SO(1,3) \otimes SU(2) \otimes U(1) \otimes SU(3)$
generators, that are the generators of standard model plus gravity.

$$\w = \left( \begin{array}[c]{cccccc}
 \vec{y}   & 0          & 0          & 0          & 0          & 0          \\
 0         & \vec{a}_1  & f+\vec{f}  & 0          & 0          & 0          \\
 0         & -f+\vec{f} & \vec{a}_2  & 0          & 0          & 0          \\
 0         & 0          & 0          & \vec{a}_3  & k+\vec{k}  & t+\vec{t}  \\
 0         & 0          & 0          & -k+\vec{k} & \vec{a}_4  & v+\vec{v}  \\
 0         & 0          & 0          & -t+\vec{t} & -v+\vec{v} & \vec{a}_5  \\
\end{array} \right) $$

\noindent We'll show in a moment that all standard model fields transform under this subgroup in the adjoint
representation. In this way themselves are elements of $Sp(12,\mathbf{C})$ algebra, explicitly:

$$\psi = \psi^1 + I\psi^2 = \left( \begin{array}[c]{cccccc}
 0              & e            & -\nu       & d^c_{R}        & d^c_{G}          & d^c_{B}   \\
 -e^*        & 0            & e^c        & -u_{R}         & -u_{G}           & -u_{B}    \\
 \nu^*       & -e^{c*}   & 0          & -d_{R}         & -d_{G}           & -d_{B}    \\
 -d^{c*}_R   & u^*_R     & d^*_R   & 0              & u^c_{B}          & -u^c_{G}  \\
 -d^{c*}_G   & u^*_G     & d^*_G   & -u^{c*}_{B} & 0                & u^c_{R}   \\
 -d^{c*}_B   & u^*_B     & d^*_B   & u^{c*}_{G}  & -u^{c*}_{R}   & 0         \\
\end{array} \right) $$

\noindent We have used the convention of Georgi - Glashow model, where the basic fields of $\psi^1$
are all left and the basic fields of $I\psi^2$ are all right. We have indicated with $^c$ the
charge conjugation. Moreover, in our formalism, $\psi^1$ and $\psi^2$ are pure quaternionic fields.
The subscripts $R,G,B$ indicates the color charge for the strong interacting
particles (R=red, G=green, B=blue).

In Georgi - Glashow model the fermionic fields are divided in two families. The first one transforms
in the representation $\bar{5}$ of $SU(5)$ (the fundamental representation). It is exactly the array
$(\w^{1j})$ in the matrix above, with $j = 2,3,4,5,6$. This array transforms in fact in the fundamental
representation for transformations in every $SU(5) \subset Sp(12,C)$ which act on indices values
$2 \div 6$.

The second family transforms in the representation $10$ of $SU(5)$ (the skew symmetric representation).
Unfortunately it isn't the sub matrix $(\w^{ij})$ with $i,j = 2,3,4,5,6$. This is in fact the skew adjoint
representation of $Sp(10,C)$, which is skew hermitian and not skew symmetric.

Do not lose heart. We'll see in a moment that such adjoint representation is a quaternionic combination
of three skew symmetric representations, one for every fermionic family. This concept could appears cumbersome,
but it will be clear along the following calculations.

\begin{theorem} The skew adjoint representation of $Sp(m)$ is a qua\-ter\-nio\-nic combination of
three skew symmetric representations of $U(m)$ plus a real skew symmetric representation (which is also
skew hermitian).
\end{theorem}

\begin{proof}
Consider a fermionic matrix $\psi$ which transforms in the adjoint representation
of $Sp(m)$:

\be \psi \ra U\psi U^\dag \label{law} \ee
Take then a matrix $\psi'$ with $\psi' k =\psi$. Its transformation law
under $U(m)$ is easily derived when this group is
constructed by using imaginary unit $i$ or $j$. This means

$$U(m) \ni U = exp(i \a_r \Sigma^r) \quad \a \in \mathbf{R}; \quad r=1,2,3,$$
$$\text{with $\Sigma$ generators of $U(m)$ whose complex entries have $i$ as imaginary unit,}$$
\begin{center}
or
\end{center}
$$U(m) \ni U = exp(j \a_r \Sigma^r) \quad \a \in \mathbf{R}; \quad r=1,2,3,$$
$$\text{with $\Sigma$ generators of $U(m)$ whose complex entries have $j$ as imaginary unit.}$$

\noindent We substitute $\psi$ with $\psi' k$ inside (\ref{law}):

$$\psi' k \ra U \psi' k U^\dag = U \psi' U^T k .$$
Here we have used the relation $k \lambda = \lambda^* k$ for $\lambda
\in \mathbf{H}$ without $k$ component. We see that $\psi'$ transforms
in the skew symmetric representation:

$$\psi' \ra U \psi' U^T$$
We obtain a complex matrix $\psi'$ (with $i$ as imaginary unit) when
$\psi$ has the form $Ak+Bj$ with $A,B$ real matrices. Indeed:

$$\psi' = - \psi k = -Akk-Bjk = A - Bi$$
Sending $\psi$ in $\psi^*$ we bring $\psi'$ to $-\psi'$ and so we satisfy the skew symmetry.
Finally we can always write

$$\psi = \psi_0 + \psi_1 k + \psi_2 i + \psi_3 j$$
In this decomposition, $\psi_1, \psi_2, \psi_3$ are complex matrices with complex unit
respectively $i, j, k$. Explicitly:

\ba \psi_1 &=& \phi_1 - i\xi_1 \qquad = \phi_1^1 - i\xi_1^1 + I(\phi_1^2 - i\xi_1^2)\nonumber \\
\psi_2 &=& \phi_2 - j\xi_2  \qquad = \phi_2^1 - j\xi_2^1 + I(\phi_2^2 - j\xi_2^2) \nonumber \\
\psi_3 &=& \phi_3 - k\xi_3  \qquad = \phi_3^1 - k\xi_3^1 + I(\phi_3^2 - k\xi_3^2).\nonumber \ea

\noindent Here all $\phi^1$, $\phi^2$, $\xi^1$, $\xi^2$ are real fields.
In this way, any $\psi_{1,2,3}$ transforms in the skew symmetric representation
of $U(m)$ when this group is built by the correspondent imaginary
unit ($i$ for $\psi_1$, $j$ for $\psi_2$ and $k$ for $\psi_3$). Hence they
define the famous three fermionic families plus a real skew symmetric field $\psi_0$. \textbf{CVD}
\end{proof}

The interaction Lagrangian can be defined as follows (with $\na = e^\mu \na_\mu$):

\ba tr(\psi^{\dag} \na \psi) &=& tr(k^* \psi^{\dag}_1 \na \psi_1 k) + tr(i^* \psi^{\dag}_2 \na \psi_2 i) +tr(j^* \psi^{\dag}_3 \na \psi_3 j) \nonumber \\
&& - tr(i^* \phi_2^\dag \na \xi_3 i) - tr(j^* \phi_3^\dag \na \xi_1 j) - tr(k^* \phi_1^\dag \na \xi_2 k) \nonumber \\
&&- tr(\psi_0^\dag \na \psi_0) \nonumber \\
 &=& tr(\psi^{\dag}_1 \na \psi_1 kk^*) + tr(\psi^{\dag}_2 \na \psi_2 ii^*) +tr(\psi^{\dag}_3 \na \psi_3 jj^*) \nonumber \\
&& - tr(\phi_2^\dag \na \xi_3 ii^*) - tr(\phi_3^\dag \na \xi_1 jj^*) - tr(\phi_1^\dag \na \xi_2 kk^*) \nonumber \\
&& - tr(\psi_0^\dag \na \psi_0) \nonumber \\
&=& tr(\psi^{\dag}_1 \na \psi_1) + tr(\psi^{\dag}_2 \na \psi_2) +tr(\psi^{\dag}_3 \na \psi_3) \nonumber \\
&& - tr(\phi_2^\dag \na \xi_3) - tr(\phi_3^\dag \na \xi_1) - tr(\phi_1^\dag \na \xi_2) \nonumber \\
&&- tr(\psi_0^\dag \na \psi_0) \label{deco}\ea

\noindent Every term $L$ in the lagrangian is intended to be integrated over $Sp(1)$:

$$L \Longrightarrow \int_{Sp(1)} dg\, g L g^{-1}.$$
In such a way, the only terms which survive are $I$-complex.
The third last line in (\ref{deco}) regroups the fermionic terms of Georgi-Glashow model for three
families in representation $10$. If we restrict $\na$ to $SU(5)$, it can be written as

$$tr\left( \left(\begin{array}{ccc} \psi_1^* & \psi_2^* & \psi_3^* \end{array}\right) \na \left(\begin{array}{c} \psi_1 \\ \psi_2 \\ \psi_3 \end{array}\right)\right) $$

\noindent where every $\psi_n$ is now constructed with $i$ as imaginary unit. This term is manifestly
invariant under global $SU(3)$ (or $SU(3) \otimes SU(3)$ if we consider also the $I$-component).
However this flavour symmetry is soon broken by mixed terms in the second last line of (\ref{deco}).
These terms give a reason to CKM and PMNS matrices which appear in the Standard Model.

In this formalism, given $\w \in su(3)\otimes su(2) \otimes u(1)$, the transformation $\d\psi = [\w,\psi]$
corresponds to the usual transformation $\d\psi = \w\psi$ in the standard model formalism.

Fields in different families are related by transformations in $Sp(1) \approx SU(2)$, ie by rotations
in the three dimensional space with base vectors $i,j,k$. Generators of $Sp(1)$ are

$$\w = \fr {\vec{y}}{6} \left( \begin{array}[c]{cccccc}
 1         & 0          & 0          & 0          & 0          & 0          \\
 0         & 1          & 0          & 0          & 0          & 0          \\
 0         & 0          & 1          & 0          & 0          & 0          \\
 0         & 0          & 0          & 1          & 0          & 0          \\
 0         & 0          & 0          & 0          & 1          & 0          \\
 0         & 0          & 0          & 0          & 0          & 1          \\
\end{array} \right) .$$

\noindent with $\vec{y} \in Im \,\mathbf{H}$.

Their diagonal form suggests an identification between this group and the gravitational
group $SU(2)^{\subset SO(1,3)}$. If the two groups coincided, all fields would transform
correctly under $SU(2)^{\subset SO(1,3)}$. By extending this group to the entire $SO(1,3)$,
we see that boosts exchange left fields with right fields.

It's remarkable that three families have to exist also for bosonic particles (photon, $W^\pm$, $Z$, gluons)
although they are probably in\-di\-stin\-gui\-sha\-ble. Note also that fields
$\psi$ appearing here don't match exactly with fermionic fields of
Standard model. The relation with these is however very simple.
Using the correspondence between hyperions and $\g\g$, the fields $\psi$
acquire two extra indices (row and column indices in $\g\g$):

$$\psi \longrightarrow \psi_{AB}$$

\noindent The standard Dirac fields have 4 components $\psi^C$ given by

$$\psi^{AB} = W^{ABC} \psi^{C}$$
where $W^{ABC}$ is any constant object which satisfies
\ba W^{*ABC} W^{ABD} = 1^{CD} &&\Longrightarrow \quad \psi^{*AB} \psi^{AB} = \psi^{*C} \psi^C \nonumber \\
    W^{*ABC} \g\g^{BF} W^{AFD} = \g\g^{CD} && \Longrightarrow \quad \psi^{*AB} \g\g^{BF} \psi^{AF} = \psi^{*C} \g\g^{CD} \psi^D \nonumber \ea

\section{Fermions from a vector superfield}
\label{superfield}
In this section we show that all fermionic and bosonic fields can be joined in a unique
superfield. This procedure doesn't need new exotic particles as \emph{squarks} or \emph{fotino}; conversely it
predicts the existence of right and sterile neutrinos. We start by introducing I-complex
grassmannian coordinates $\theta = \theta^1 + I\theta^2$ and $\bar{\theta} = \theta^1 - I\theta^2$
with obvious fundamental products:

\ba \theta\theta &=& \theta^1 \theta^1 + \theta^1 I \theta^2 + I \theta^2 \theta^1 - \theta^2\theta^2 = 0 + I \theta^1 \theta^2 - I \theta^1 \theta^2 - 0 = 0 \nonumber \\
 \bar{\theta}\bar{\theta} &=& \theta^1 \theta^1 - \theta^1 I \theta^2 - I \theta^2 \theta^1 - \theta^2\theta^2 = 0 - I \theta^1 \theta^2 + I \theta^1 \theta^2 - 0 = 0 \nonumber \\
 \theta\bar{\theta} &=& \theta^1 \theta^1 - \theta^1 I \theta^2 + I \theta^2 \theta^1 + \theta^2\theta^2 = - I \theta^1 \theta^2 - I \theta^1 \theta^2 = -2 I \theta^1 \theta^2 \nonumber \ea

\noindent Accordingly, there will exist grassmannian derivatives $\pa_g$ and $\bar{\pa}_g$ with $\pa_g \theta = \bar{\pa}_g \bar{\theta} =1$ and $\pa_g \bar{\theta} = \bar{\pa}_g \theta =0$. At this point we can define a new supersymmetric algebra as follows:

\ba Q = \pa_g - e^\mu \bar{\theta} \pa_\mu \qquad\quad\!\! &;&\quad\qquad\!\! [Q, P_\nu] = \bar{\theta} (\pa_\nu e^\mu)P_\mu \nonumber \\
\bar{Q} = \bar{\pa}_g - \theta \bar{e}^{\dag \nu} \pa_\nu \qquad\quad\!\! &;&\quad\qquad\!\!
\{Q, \bar Q \} = 2I{\Sigma}^{\mu} P_\mu \nonumber \ea
$$ 2 e^\nu_H = e^\nu + \overline{e}^{\dag\nu} \qquad\quad\,\,\,\,\,\, P_\mu = - I\pa_\mu $$
$$2{\Sigma}^{\mu} = 2 e^\mu_H + \theta \bar{\theta} \left[e^\rho \pa_\rho \bar{e}^{\dag\mu} - \bar{e}^{\dag\rho}\pa_\rho e^\mu\right]  $$

\noindent The most general superfield is then

$$V(x,\theta,\bar{\theta}) = e^\mu (x) A_\mu (x) + \theta \psi(x) + \bar{\chi}(x)\bar{\theta} + \theta \bar{\theta} F(x) .$$

\noindent To obtain an irreducible representation of SUSY algebra we introduce a covariant derivative $\bar{D}$ which commutes both with $Q$ and $\bar{Q}$:

$$\bar{D} = \bar{\pa}_g + \theta e^{\nu} \pa_\nu$$

\noindent In terms of shifted coordinates $y^\mu = x^\mu + e^\mu \theta \bar\theta$, the action of $\bar{D}$ simplifies considerably:

$$\bar{D} V(y,\theta,\bar{\theta}) = \bar{\pa}_g V(y,\theta,\bar{\theta})$$

\noindent In this way we can define a supersymmetric chiral field by imposing $\bar{\pa}_g V(y,\theta,\bar{\theta}) = 0$, whose solution is clearly

$$V \equiv V(y,\theta) = e^\mu (y) A_\mu (y) + \theta \psi(y).$$
In the original coordinates this gives

$$V(x,\theta, \bar{\theta}) = e^\mu (x) A_\mu (x) + \theta \psi(x) + \theta \bar{\theta} e^\nu \pa_\nu (e^\mu A_\mu)$$

\noindent Infinitesimal SUSY transformations induced by $\epsilon Q + \bar{\epsilon} \bar{Q}$ are easily computed:

\ba \d \psi &=& -2\bar{\epsilon} e_H^\nu \pa_\nu  (e^\mu A_\mu) \nonumber \\
\d A_\mu &=& \epsilon e_\mu \psi \ea

\noindent From $V$ we can construct a generalized covariant derivative
by substituting $A_\mu$ with $\pa_\mu + A_\mu$ and $\psi$ with $\pa_g +\psi$:

$$\na = e^\mu (\pa_\mu + A_\nu) + \theta (\pa_g + \psi) + \theta \bar{\theta} e^\nu \pa_\nu [e^\mu (\pa_\mu + A_\mu)]$$

\noindent It can be useful to introduce derivatives $\pa_{g\mu}$ and fields
$\psi_\mu$ in representation $\left(1 \otimes \fr 12 \right) = \left(\fr 12 \oplus \fr 32\right)$
with properties $e^\mu \pa_{g\mu} = \pa_g$ and $e^\mu \psi_\mu = \psi$. In this way

$$\na = e^\mu \na_\mu = e^\mu \left\{ \pa_\mu + A_\mu + \theta (\pa_{g\mu} + \psi_\mu) + \theta \bar{\theta} \pa_\mu [e^\nu (\pa_\nu + A_\nu)]\right\}$$

\noindent New SUSY transformation laws emerge for $\pa_\mu$ and $\pa_g$:

\ba \d \pa_g &=& -2\bar{\epsilon} e_H^\nu (\pa_\nu  e^\mu \pa_\mu + e^\mu A_\mu \pa_\nu) \nonumber \\
\d \pa_\mu &=& \epsilon e_\mu \pa_g \ea

\noindent By composing quadratic and quartic powers of $\na$ and $\overline{\na}^\dag$, you can extract
all terms which appear in Standard Model, plus Hilbert-Einstein and Gauss-Bonnet terms.

It's remarkable that standard fermionic fields take the role of gauginos for
standard gauge fields. In this way the right up quarks are gauginos for gluons,
while right electrons are gauginos for $W$ bosons. Clearly this is permitted because
fermions and bosons transform in the same representation of
$Sp(12,\mathbf{C})$. In such a way our theory includes SUSY $N = 1$ with no need
for new unknown fermionic or scalar particles, apart from one exception.

SUSY predicts the existence of a new colored fermionic sextuplet which sits
on diagonal in $\psi$. Inside it we can include a conjugate neutrino ($\nu^c$),
a sterile neutrino ($N$) and a conjugate sterile neutrino ($N^c$). Explicitly

$$\psi = \left( \begin{array}[c]{cccccc}
N & 0 & 0 & 0 & 0 & 0 \\
0 & \nu^c & 0 & 0 & 0 & 0 \\
0 & 0 & \nu^c & 0 & 0 & 0 \\
0 & 0 & 0 & N^c & 0 & 0 \\
0 & 0 & 0 & 0 & N^c & 0 \\
0 & 0 & 0 & 0 & 0 & N^c \end{array}
\right).$$

\noindent This field commutes with any gauge field in $U(1) \otimes SU(2) \otimes SU(3)$
and so it hasn't electromagnetic, weak or strong interactions. Moreover it gives a Dirac mass to
neutrinos via the term

$$tr\,(\bar{\psi}^\dag e^\mu A_\mu \psi) = \bar{\psi}^{\dag ij} e^\mu A_\mu^{kl} \psi^{mn} f^{(ij)(kl)(mn)}.$$
Here $f^{(ij)(kl)(mn)}$ are structure constants for $SU(6)$ and masses for neutrinos are eigenvalues of $<e^\mu A_\mu >$.

\section{The octonions hypothesis}

We indicate by $x$ a generic \lq\lq number'' equipped with $7$ imaginary components.
This number can be considered both hyperionic and octonionic, where octonions
are defined as here $\,\Rightarrow \, http: //en.wikipedia.org/$ $wiki/Octonion $. Explicitly:

$$x = a + i_1 b + i_2 c + i_3 d + I w + i_1 I s + i_2 I g + i_3 I t$$
$$a,b,c,d,w,s,g,t \in \mathbf{R}.$$
Note that we have written $i_1,i_2,i_3$ in place of $i,j,k$.
The tabular summarizes the differences between hyperionic and octonionic case:

\begin{center}
\begin{tabular}{|p{5cm}|p{5cm}|}
Hyperions ($n \neq m$) & Octonions ($n \neq m$)\\
\hline
$i_n i_m = \e^{nmq} i_q $ & $i_n i_m = \e^{nmq} i_q$ \\
$I i_n = i_n I$ & $I i_n = -i_n I$ \\
$I^2 = -1$ & $I^2 = -1$ \\
$\overline{(i_n I)}^\dag = -i_n I$ & $(i_n I)^\dag = -i_n I$ \\
$i_n (i_m I) = (i_n i_m) I$ & $i_n (i_m I) = -(i_n i_m) I$ \\
$i_n (i_m I) = -(i_m I)i_n $ & $i_n (i_m I) = -(i_m I) i_n $ \\
$i_n (i_n I) = (i_n i_n) I = -I$ & $i_n (i_n I) = (i_n i_n) I = -I$ \\
$(i_n I)(i_m I) = - i_n i_m$ & $(i_n I)(i_m I) = -i_n i_m$ \\
$(i_n I)^2 = 1$ & $(i_n I)^2 = -1$
\end{tabular}
\end{center}

\noindent It's considerable that all differences (also non-associativity of octonions)
arise by imposing $I i_n = - i_n I$ without changing $i_n (i_m I) = - (i_m I) i_n$.
Let consider the following octonionic field:

$$\Phi = e^\mu (x) A_\mu + i_1 \l^1_\a (\theta) \psi^\a_{1} + i_2 \lambda^2_\b (\theta) \psi^\b_{2} + i_3 \l^3_\g (\theta) \psi^\g_{3}$$

$$\Phi = E^A W_A \qquad\quad E^A =\left( \begin{array}{c} \l^1_\a \\ \l^2_\b \\ \l^3_\g \\ e^\mu \end{array} \right) \qquad W_A =\left(  i_1 \psi^\a_1 \, ; \, i_2 \psi^\b_2 \, ; \, i_3 \psi^\g_3 \, ; \, A_\mu \right)$$

$$ A_\mu = Re\, A_\mu + I Im\,A_\mu \qquad\quad\qquad \psi^\a_{n} = Re\,\psi^\a_{n} + I Im\,\psi^\a_{n} $$

\noindent Here $\psi^\a_{n}$ and $\theta^\a$ are usual Weyl spinors with two components ($\a =1,2$).
$\l^n_\a$ are octonionic functions of $\theta^\a$. At this point we can define a generalized covariant
derivative and a generalized metric:

$$\na_A =\left(  \pa^\a_{g1} + i_1 \psi^\a_1 \, ; \, \pa^\b_{g2} + i_2 \psi^\b_2 \, ; \, \pa^\g_{g3} + i_3 \psi^\g_3 \, ; \, \pa_\mu + A_\mu \right)$$

$$H^{AB} = Re\,(E^A E^B)$$

\noindent We conjecture that actions of both Standard Model and General Relativity are comprised inside an
action of the following type:

$$S = \int \sqrt H\, d^{10} X \,(E^2)^{AB} (\na^2)_{AB} + g \int \sqrt H \, d^{10} X \,  (E^4)^{ABCD} (\na^4)_{ABCD} \qquad\quad g \in \mathbf{R}$$

$$X^A = \left( \begin{array}{c} \theta^1_\a \\ \theta^2_\b \\ \theta^3_\g \\ x^\mu \end{array} \right) \qquad H = det(H^{AB})$$

\noindent In a such action, super-symmetry is replaced by covariance under generalized coordinates transformations:

\ba
X &\ra& X'(X) \nonumber \\
x &\ra& x'(x, \theta) \nonumber \\
\theta &\ra& \theta' (x, \theta) \nonumber \ea

\noindent Interpret now $A_\mu$ and $\psi_n$ as $6 \times 6$ complex matrices. Moreover we want $A_\mu$ being
gauge field for $SU(6)$ and so it will be skew-hermitian. Finally we choose $\psi_n$
skew-symmetric in such a way to have $i_n \psi_n$ skew-hermitian. Hence $W$ will result
skew-hermitian too.

\ba && \Phi^{ij} = e^\mu (x) \left(A^{ij}_\mu + \d^{ij}\overset{G}{A_\mu}\right)  + i_1 \l^1_\a (\theta) (\psi^\a_{1})^{ij} + i_2 \lambda^2_\b (\theta) (\psi^{\b}_{2})^{ij} + i_3 \l^3_\g (\theta) (\psi^{\g}_{3})^{ij} \nonumber \\
&& \overset{G}{A_\mu} = i (a_\mu + I a'_\mu) + j(b_\mu + I b'_\mu) + k (c_\mu + I c'_\mu) \nonumber \\
&&\qquad\qquad\qquad\qquad\qquad\qquad\qquad\qquad\qquad\qquad\qquad a_\mu,a'_\mu,b_\mu, b'_\mu, c_\mu,c'_\mu \in \mathbf{R} \nonumber \\
&& (\psi_n^\a)^i_{\pt i} = 0 \nonumber \ea

\noindent Here we have considered as fermionic the imaginary components proportional to $i_n$ and $i_n I$, except for the trace. This last is taken bosonic and obviously it gives the gravitational field. Moreover we can take $\l^n_\a(\theta) = \theta^n_\a$ as the most natural metric. At this point we can study the action of $SU(6)$ on $W$, where $SU(6)$ is built with $I$ as imaginary unit:

$$W_A \ra U^\dag W_A U - U^\dag \pa_A U \qquad\quad U\in SU(6)$$
$$A_\mu \ra U^\dag A_\mu U - U^\dag \pa_\mu U$$
$$i_n \psi^\a_n \ra U^\dag i_n \psi^\a_n U - U^\dag \pa^\a_{g n} U$$
$$\Downarrow$$
$$\psi^\a_n \ra U^T \psi^\a_n U$$

\noindent In the last step we have used $\pa^\a_{g n} U = 0$ and $i_n I = - I i_n$.
You see that fermionic fields still transform in the skew-symmetric representation
and so they fit easily with standard fermionic fields.

However, skew-hermitian matrices with entries in $\mathbf{O}$ don't define
a Lie Algebra, due to non-associativity of octonions. Conversely they define a \textbf{ternary
algebra}, whose corresponding gauge theory is well discussed in \cite{ternary}.
Their exponentiated version is now the unitary \lq\lq quasi''-group $T6$, where the word \lq\lq quasi''
underlines the lack of associativity, which is an ordinary request in the usual definition of group.
In such theories the field strength results:

$$R_{AB} = \pa_A W_B - \pa_B W_A + [W_A, W_B, g]$$

\noindent where $g$ is an auxiliary octonionic field and the 3-bracket is defined as follows:

$$[u,v,x] = D_{u,v} x = \fr 12 \left( u(vx) - v(ux) + (xv)u - (xu)v + u(xv) - (ux)v \right).$$

\noindent Gauge transformations of $W$ and $g$ are given in \cite{ternary}, while $R$ transforms
homogeneously as expected. Note that, if $u,v,x$ belong to an associative algebra, then $[u,v,x] = \fr 12 [[u,v],x]$. To satisfy all requests of your model, we have to find an auxiliary field $g$ such that:

$$Re\, [A_\mu, A_\nu, g][A_\rho, A_\s, g] e^\mu e^\nu e^\rho e^\s = Re\, [A_\mu, A_\nu][A_\rho, A_\s]e^\mu e^\nu e^\rho e^\s$$
$$Re\, [\overset{G}{A_\mu}, \overset{G}{A_\nu}, g] e^\mu e^\nu = Re\,[\overset{G}{A_\mu}, \overset{G}{A_\nu}] e^\mu e^\nu$$
where in the left side we consider $i_n, i_n I$ octonionic, while in the right side we consider them hyperionic.
Clearly much work remains to do, but is clear that, moving from Hyperions to Octonions,
we lost all the oddities of previous sections, namely the tripling of $SU(6)$
gauge fields and the existence of a strange real field $\psi_0$. The resulting
symmetry group will be $T6$ in place of $Sp(12, C)$.

\subsection{Extension of Ricci scalar}

Note that such framework provides fermionic contributes to Ricci scalar. Explicitly:

$$R = R^{BOS} +\int d^2\theta\,\left(- e^{\dag \mu} [\na_\mu, \na_\a] \l^\a + \l^{\dag \b} [\na_\b^\dag, \na_\nu] e^\nu + \l^{\dag \b} \{\na^\dag_\b, \na_\a\} \l^\a \right)$$

\noindent Varying the last term with respect to $\psi$ we obtain

$$\na^\dag_\b \l^\a = (\pa^\dag_\b - i\psi_\b^\dag)\l^\a = 0.$$

\noindent A good distributional solution is then

$$\l^a = \int d^2 \xi \,\xi^\a e^{i\theta^{\dag\t} \psi^\dag_\tau}.$$

\noindent Considering that $[\na_\b^\dag, \na_\nu] = -i[\psi_\b^\dag, \na_\nu] = i[\na_\nu, \psi_\b^\dag]$, the second last term becomes

\ba (2^{ND} \,\,\, LAST) &=& i \int d^2 \xi^\dag \, d^2 \theta \, \xi^{\dag\b} e^{-i \theta^\tau \psi_\tau} \left[\na_\nu, \psi_\b^\dag\right] e^\nu \label{rela} \\
 &=& i \int d^2 \xi^\dag \, d^2 \theta \, \xi^{\dag\b} \left(1 -i \theta^\tau \psi_\tau -\fr 12 \theta^\tau \psi_\tau \theta^\g \psi_\g \right) \left[\na_\nu, \psi_\b^\dag \right] e^\nu \nonumber\ea

\noindent Apply now the standard formulas for grassmannian integrals, ie $\theta^\tau \theta^\g = \fr 12 \e^{\tau\g} \theta^2$, $\int d^2 \theta \, \theta^2 = 2$, $\int d^2 \theta = 0$ and $\int d^2 \theta \, \xi^{\dag\b}\theta^\tau = \e^{\b\t} \d^2 (\xi^\dag - \theta)$. Relation (\ref{rela}) simplifies in

\ba (2^{ND} \,\,\, LAST) &=& \int d^2 \xi^\dag \, \left( \d^2 (\xi^\dag - \theta) \psi^\b +\fr i2 \xi^{\dag\b} \psi^\g \psi_\g \right) \left[\na_\nu, \psi_\b^\dag \right] e^\nu \nonumber \\
 &=& \psi^\b [\na_\nu, \psi_\b^\dag] e^\nu , \ea

\noindent where we have used $\int d^2 \xi^\dag \, \xi^{\dag\b} = 0$. In this way we have obtained the kinetic term for fermions directly from Ricci scalar. However this works exactly if $\l^\a$ is a $6 \times 6$ matrix of spinor, and not a simple spinor. It's notable that the same process can be utilized to obtain kinetic terms for gauge fields by starting with an $e^\mu$ intended as a $6 \times 6$ octonionic matrix of vectors:

$$e^\mu (x) = \eta^\mu e^{- \int^x dx^{\prime \nu} A_\nu (x')} \qquad\qquad \eta^{\mu\nu} = Re\,(\eta^\mu \eta^{\dag\nu}).$$

\section{Antigravity}
\label{antig}
\begin{spacing}{1.5}
The kinetic piece in lagrangian (\ref{quartic}) includes the
following term which mixes gravity with electromagnetism:
\end{spacing}
\normalsize
$$ -\fr 14 f^{(G)(EM1)(EM2)} A^{(G)}_\mu A^{(EM1)}_\nu \bigg(
F^{(EM2)\mu\nu} + \a f^{(EM3)(EM1)(EM2)} A^{(EM3)\mu} A^{(EM1)\nu}
\bigg) $$
\normalsize
\be \pt \label{prima}\ee
Remember that AFT includes three indistinguishable e\-lec\-tro-ma\-gne\-tic
fields, with non-trivial commutators. In this way $A^{(G)}$ is the
gravitational gauge field, $A^{(EMn)}$ is the n-th electromagnetic
field and $\a$ is the fine structure constant. In the realistic case of
null torsion, the gravitational gauge field can be rewritten in
function of the tetrad field:

$$A_\mu^{(G)bc} = \fr 12 e^{\nu [b} \pa_{[\mu} e^{c]}_{\nu]} + \fr 14
e_{\mu d} e^{\nu b} e^{\sigma c} \pa_{[\sigma} e^d_{\nu]}$$

\noindent From now we take a low energy limit so defined: $e_{ii} = 1$ with $i=1,2,3$,
$e_{00} = \theta(x)$ and $\pa_0 \theta(x) =0$. Varying with respect to $e$
we obtain:

$$\fr {\d A_\mu^{(G)bc}}{\d e^s_\tau} = \fr 12 e^{\nu [b} \d^{c]}_s
\d^\tau_{[\nu} \pa_{\mu]} + \fr 14 e_{\mu s} e^{\nu b} e^{\sigma c}
\d^\tau_{[\nu} \pa_{\sigma]}$$

$$\fr {\d A_\mu^{(G)bc}}{\d g_{\w\tau}} = 2e^{\w s} \fr {\d A_\mu^{(G)bc}}{\d e^s_\tau}
= e^{\w [c} e^{b] \nu} \d^\tau_{[\nu} \pa_{\mu]} + \fr 12 \d^\w_{\mu} e^{\nu b} e^{\sigma c}
\d^\tau_{[\nu} \pa_{\sigma]}$$
The component with $c = \w =\tau = 0$ and $b \neq 0$ results:

$$\fr {\d A_\mu^{(G)b0}}{\d g_{00}} = -\theta^{-1} \d^0_\mu \pa_b
- \fr 12 \theta^{-1} \d^0_\mu \pa_b = -\fr 3{2\theta} \d^0_\mu \pa_b$$

$$A^{(EM)\rho}A^{(EM)}_\rho A^{(EM)\mu}\fr {\d A_\mu^{(G)b0}}{\d g_{00}} = \fr 3{2\theta} \pa_b A^{(EM)0} A^{(EM)\rho}A^{(EM)}_\rho$$
The minus sign has disappeared because we have reversed the derivative.
The variation of quartic term in (\ref{prima}) with respect to $\d g_{00}$
is then given by

$$-\fr \a 4 \cdot \fr 3{2\theta} \pa_b f^b A^{(EM)0} A^{(EM)\rho}A^{(EM)}_\rho =
-\pa_b f^b \fr {3\a} {8\theta} V(\theta^2 V^2 - A^2)$$
$$f^b = \sum_{cade} f^{(bo)ca} f^{dea} \approx 4\fr {x^b}{r}.$$
Here we have indicated with $V$ the electric potential and with $A$ the
magnetic vector potential. The sum inside $f$ is over the three electromagnetic
fields.

It's so clear that varying the complete action with respect to $g_{\mu\nu}$
we obtain a new term for Einstein equations. In the Newtonian limit we can
substitute $g_{00} = -(1-2\phi)$ and $R_{00} - (1/2)Rg_{00} = \na^2 \phi$
where $\phi$ is the newtonian potential. Hence:

\ba 2 \na^2 \phi &\approx& 8\pi T^{00} = 8\pi \fr{-2}{\sqrt {-g}} \fr {\d \sqrt {-g} L_{matt}}{\d g_{00}} \nonumber \\
            &\approx& \pa_b \fr {x^b}{r} 24\pi \a V(\theta V^2 - \theta^{-1} A^2) \ea
For radial potential we have

$$\pa_b \phi = \fr {x^b}{r} \pa_r \phi .$$
In such case

$$C_G = \pa_r \phi \approx 12\pi\a V(\theta V^2 -\theta^{-1} A^2)$$
Now we insert the appropriate universal constants and approximate $\theta$  with $1$:

\be C_G \approx 12\pi \a \fr{(G\e_0)^{3/2}}{c^4 L_p} V(V^2 - c^2 A^2) = k V(V^2 - c^2 A^2) \label{ultima} \ee
Here $L_p$ is the Planck length, equal to $\sqrt {\hbar G/c^3}$. The multiplicative constant is

$$ k = \fr{12\pi}{137}\cdot \fr{(6,67\cdot 10^{-11}\cdot 8,85\cdot 10^{-12})^{3/2}}{(3\cdot 10^8)^4 \cdot(1,62\cdot 10^{-35})} = 30,27\cdot 10^{-33} \,\left(\fr{C^3 s^4}{Kg^3 m^5}\right).$$
This means that for having a weight variation (on Earth) of about $10\%$
($\Delta C_G =1$) we need an electrical potential of $10^{11}$ Volts.
These are $100$ billions of Volts. For $V = Q/r$ and $A=0$ we have:

$$C_G = \fr {k}{(4\pi\e_0)^3}\cdot \fr {Q^3}{r^3} = 2,198 \cdot 10^{-2} \left(\fr {m^4}{s^2 C^3}\right)\fr {Q^3}{r^3}$$
Note that the sign of $C_G$ is the sign of $Q$ and then we obtain
antigravity for negative $Q$. We associate to this interaction an
equivalent mass $m$, substituting $C_G = Gm/{r^2}$. We have

$$m = \fr k G V^3 r^2 = \fr {k}{G(4\pi \e_0)^3}\fr {Q^3} r = 3,293\cdot 10^8 \left(\fr {Kg\, m}{C^3}\right) \fr {Q^3}{r}$$
which is a negative mass for negative $Q$. Negative mass implies negative
energy via the relation $E =mc^2$.
Intuitively, if we search a similar relation for gravi-magnetic field
(which is $\na \times (g^{0i})$, $i=1,2,3$), we should find the same
formula (\ref{ultima}) with an exchange between $V$ and $cA$.

We calculate now at what distance the gravitational attraction between two
protons is equal to their electromagnetic repulsion.

$$G\fr {m^2}{r^2} = \fr {k^2}{G^2 (4\pi\e_0)^6} \fr {Q_p^6}{r^4} = \fr 1 {4\pi \e_0} \fr {Q_p^2}{r^2}$$

$$\fr {k^2 Q_p^4}{G^2 (4\pi\e_0)^5} = r^2$$

$$\Longrightarrow r^2 = 79,49 \cdot 10^{-70} m^2 \Longrightarrow r = 8,916 \cdot 10^{-35} m = 5,516\, L_p$$
Note that we are $20$ orders of magnitude under the range of strong force
and $23$ orders of magnitude under the range of weak force. In this way
the gravitational force doesn't affect the making of nucleus and nucleons.

\section{Conclusion}

In the course of paper we have demonstrated that a satisfactory
gauge theory exists which includes all the four forces.
However, if we try to quantize the theory, we encounter the well
known renormalization problems for diagrams which involve the
tetrad field $e^\mu$. The complete theory, exposed in \cite{Arrangement},
overcomes this trouble by quantizing theory before the choice of
a fixed spin-network, in such a way that $e^\mu$ has still to born.

Another possibility is suggested by the analogy between $e^\mu$ and $\theta$
in the superfield expansion, united to the role of $e^\mu$ inside the
generalized coordinate $y^\mu$. In fact we can consider $e^\mu$ as another quadruplet
of coordinates, so that the other fields become:

\ba A_\mu(x^\nu) &&\ra A_\mu (x^\nu, e^\nu) \nonumber \\
    \psi_\mu (x^\nu) &&\ra \psi(x^\nu, e^\nu) \nonumber \\
    d^4x &&\ra d^4x d^4e \ea

A conjugated momentum $p^e_\mu$ will be associated to $e^\mu$, while
in Feynmann diagrams we'll have to substitute $e^\nu$ with $-I\pa /\pa p^e_\nu$.

Back to the present work, in the last section we have seen that a
potential of $10^{11}$ Volts can induce relevant gra\-vi\-ta\-tio\-nal
effects. They are too many for notice variations in the experiments
with particles accelerators. However they sit at the border of our
technological capabilities.

We hope that a future team work shall explore this theory in detail,
deepening also the triality with strings and loop gravity, highlighted in
\cite{Arrangement}.

\end{document}